\documentclass[conference]{IEEEtran}
\IEEEoverridecommandlockouts

\usepackage{cite}
\usepackage{amsmath,amssymb,amsfonts}
\usepackage{mathtools}
\usepackage{algorithm}
\usepackage{algorithmic}
\usepackage{graphicx}
\usepackage{textcomp}
\usepackage{xcolor}
\usepackage{physics}
\usepackage{booktabs}
\usepackage{multirow}

\def\BibTeX{{\rm B\kern-.05em{\sc i\kern-.025em b}\kern-.08em
    T\kern-.1667em\lower.7ex\hbox{E}\kern-.125emX}}

\begin{document}

\title{Quantum Phase Classification of Rydberg Atom Systems Using Resource-Efficient Variational Quantum Circuits and Classical Shadows}

\author{\IEEEauthorblockN{Hemish Ahuja\IEEEauthorrefmark{1}, Samradh Bhardwaj\IEEEauthorrefmark{2}, Kirti Dhir\IEEEauthorrefmark{3}, Roman Bagdasarian\IEEEauthorrefmark{4}, and Ziwoong Jang\IEEEauthorrefmark{5}}
\IEEEauthorblockA{\IEEEauthorrefmark{1}\IEEEauthorrefmark{3}York University, Toronto, Canada}
\IEEEauthorblockA{\IEEEauthorrefmark{2}Modern School Vaishali, India}
\IEEEauthorblockA{\IEEEauthorrefmark{4}Johannes Kepler Universität Linz, Linz, Austria}
\IEEEauthorblockA{\IEEEauthorrefmark{5}Pusan National University, Busan, South Korea}}

\maketitle

\begin{abstract}
Quantum phase transitions in Rydberg atom arrays present significant opportunities for studying many-body physics, yet distinguishing between different ordered phases without explicit order parameters remains challenging. We present a resource-efficient quantum machine learning approach combining classical shadow tomography with variational quantum circuits (VQCs) for binary phase classification of Z$_2$ and Z$_3$ ordered phases. Our pipeline processes 500 randomized measurements per 51-atom chain state, reconstructs shadow operators, performs PCA dimensionality reduction (51$\rightarrow$4 features), and encodes features using angle embedding onto a 2-qubit parameterized circuit. The circuit employs RY-RZ angle encoding, strong entanglement via all-to-all CZ gates, and a minimal 2-parameter ansatz achieving depth 7. Training via simultaneous perturbation stochastic approximation (SPSA) with hinge loss converged in 120 iterations. The model successfully learned to separate the classes in the limited dataset, achieving 100\% accuracy on the training, validation, and held-out test sets. This work establishes pathways for quantum-enhanced condensed matter physics on near-term quantum devices.
\end{abstract}

\begin{IEEEkeywords}
quantum machine learning, Rydberg atoms, phase classification, variational quantum circuits, classical shadows, SPSA optimization
\end{IEEEkeywords}

\section{Introduction}

Quantum phases of matter represent distinct organizational patterns characterized by spontaneous symmetry breaking and long-range quantum correlations \cite{sachdev2011}. Understanding and identifying these phases is fundamental to condensed matter physics and quantum simulation. Recent experimental advances in programmable Rydberg atom arrays have enabled the study of exotic quantum phases with unprecedented control \cite{bernien2017,ebadi2021}, yet classifying these phases without explicit order parameters remains a significant challenge.

\subsection{Background and Motivation}

Rydberg atoms, with their strong van der Waals interactions ($V(r) \sim C_6/r^6$) and controllable Rydberg blockade effect, have emerged as a versatile platform for quantum simulation \cite{saffman2010}. In one-dimensional chains, these systems exhibit rich phase diagrams including Z$_2$ (period-2 ordering) and Z$_3$ (period-3 ordering) phases depending on the detuning $\delta$ and blockade radius $R_b$ parameters. Traditional phase classification methods rely on computing order parameters specific to each phase, requiring detailed knowledge of the system's ground state—information that is exponentially expensive to obtain via full quantum state tomography.

The challenge intensifies in realistic experimental scenarios where only limited measurements are available. Full tomography of an $N$-qubit system requires $O(4^N)$ measurements, which is infeasible for even moderate system sizes. For a 51-atom Rydberg chain, this corresponds to approximately $10^{30}$ measurements—far beyond experimental capabilities.

\subsection{Problem Statement}

Given measurement outcomes from a 51-qubit Rydberg atom chain in an unknown quantum phase, with only 500 measurements per state in randomized Pauli bases, our goal is to perform binary classification: determine whether the system is in Z$_2$ or Z$_3$ ordered phase using minimal quantum circuit resources suitable for near-term quantum hardware. The constraints include: (1) no direct access to ground state wavefunction, (2) limited measurement budget, and (3) NISQ device limitations requiring shallow, narrow circuits.

\subsection{Contributions}

This work makes the following novel contributions:

\textbf{1) Efficient Preprocessing Pipeline:} We develop a classical shadow reconstruction protocol combined with PCA dimensionality reduction and angle-based feature encoding, reducing the problem from 51 qubits to 4 encoded features while preserving phase signatures.

\textbf{2) Minimal Circuit Architecture:} We design a 2-qubit variational circuit with depth 7 and only 2 trainable parameters, incorporating strong entanglement for correlation capture while avoiding barren plateau problems.

\textbf{3) Hardware-Compatible Optimization:} We implement SPSA gradient-free optimization with hinge loss for maximum-margin classification, demonstrating rapid convergence and robustness.

\textbf{4) Perfect Classification:} We achieve 100\% test accuracy (3/3 samples), 100\% validation accuracy (3/3 samples), and 100\% training accuracy (14/14 samples), with a resource efficiency score of 0.5986.

\section{Theoretical Background}

\subsection{Rydberg Atom Systems}

The Rydberg Hamiltonian for a one-dimensional chain of $N$ atoms is given by:

\begin{equation}
H = \frac{\Omega}{2}\sum_{i=1}^{N} X_i - \delta \sum_{i=1}^{N} n_i + \sum_{i<j} \frac{\Omega R_b^6}{(a|i-j|)^6} n_i n_j
\label{eq:rydberg_hamiltonian}
\end{equation}

where $\Omega$ is the Rabi frequency driving transitions between ground $\ket{g}$ and Rydberg $\ket{r}$ states, $\delta$ is the detuning parameter, $n_i = \ket{r}\bra{r}_i$ is the Rydberg number operator, $R_b$ is the blockade radius, and $a$ is the lattice spacing.

The first term represents coherent Rabi coupling, the second term penalizes Rydberg excitations proportional to detuning, and the third term captures van der Waals interactions between Rydberg atoms. The Rydberg blockade effect prevents multiple atoms within distance $R_b$ from being simultaneously excited, fundamentally shaping the system's phase diagram.

\subsubsection{Phase Diagram and Ordering Patterns}

The system exhibits distinct quantum phases characterized by spatial ordering of Rydberg excitations:

\textbf{Z$_2$ Phase (Period-2):} Ground state approximately $\ket{\ldots rgrgrg \ldots}$ with alternating Rydberg excitations, breaking translational symmetry by 2 sites.

\textbf{Z$_3$ Phase (Period-3):} Ground state approximately $\ket{\ldots rggr ggr gg \ldots}$ with Rydberg excitations separated by 2 ground-state atoms, breaking translational symmetry by 3 sites.

Phase transitions occur by varying $\delta$ and $R_b$. Our dataset samples 20 points across this parameter space (10 Z$_2$, 10 Z$_3$), each prepared in the ground state and measured.

\subsection{Classical Shadow Tomography}

Classical shadows, introduced by Huang et al. \cite{huang2020}, enable efficient estimation of quantum state properties using logarithmically fewer measurements than full tomography.

\subsubsection{Shadow Generation Protocol}

For each measurement $t = 1, \ldots, T$:
\begin{enumerate}
\item Choose random Pauli basis: $U_t$ (measurement in $X$, $Y$, or $Z$ eigenbasis)
\item Measure in computational basis: outcome $b_t$
\item Compute shadow operator:
\begin{equation}
\sigma_t = 3 U_t^\dagger \ket{b_t}\bra{b_t} U_t - I
\label{eq:shadow_operator}
\end{equation}
\end{enumerate}

The classical shadow estimate of density matrix $\rho$ is:
\begin{equation}
\hat{\rho} = \frac{1}{T}\sum_{t=1}^{T} \frac{\sigma_t + I}{3}
\label{eq:shadow_reconstruction}
\end{equation}

This provides an unbiased estimator: $\mathbb{E}[\hat{\rho}] = \rho$.

\subsubsection{Measurement Basis Mapping}

Our data provides measurements in Pauli eigenstates: $\{g, r, +, -, +i, -i\}$ corresponding to:

\begin{align}
\ket{g} &= \begin{pmatrix} 1 \\ 0 \end{pmatrix}, \quad \ket{r} = \begin{pmatrix} 0 \\ 1 \end{pmatrix} \\
\ket{+} &= \frac{1}{\sqrt{2}}\begin{pmatrix} 1 \\ 1 \end{pmatrix}, \quad \ket{-} = \frac{1}{\sqrt{2}}\begin{pmatrix} 1 \\ -1 \end{pmatrix} \\
\ket{+i} &= \frac{1}{\sqrt{2}}\begin{pmatrix} 1 \\ i \end{pmatrix}, \quad \ket{-i} = \frac{1}{\sqrt{2}}\begin{pmatrix} 1 \\ -i \end{pmatrix}
\end{align}

These are eigenstates of Pauli $Z$, $X$, and $Y$ operators respectively. For each outcome $\psi$, we construct:
\begin{equation}
\sigma(\psi) = 3\ket{\psi}\bra{\psi} - I_2
\end{equation}

\subsubsection{Advantage Over Full Tomography}

Classical shadows require $O(\log N)$ measurements to estimate properties of $N$-qubit states with bounded error, compared to $O(4^N)$ for full tomography. For our 51-qubit system, 500 measurements suffice to capture phase-distinguishing features—a $\sim10^{27}$-fold reduction in measurement complexity.

\subsection{Variational Quantum Circuits}

Variational quantum circuits are parametrized quantum gates $U(\theta)$ that prepare quantum states $\ket{\psi(\theta)} = U(\theta)\ket{0}$ for hybrid quantum-classical optimization. The circuit comprises:

\textbf{Feature Encoding:} Maps classical data $x$ to quantum state
\textbf{Ansatz:} Parametrized gates with trainable weights $\theta$
\textbf{Measurement:} Extracts classical information (e.g., $\langle Z \rangle$)

A cost function $C(\theta)$ is minimized via classical optimizer updating $\theta$ based on quantum circuit evaluations. Key challenges include barren plateau problems (vanishing gradients in deep circuits) \cite{mcclean2018} and the expressibility-trainability tradeoff \cite{cerezo2021}.

For phase classification, VQCs can capture quantum correlations in a sample-efficient manner, potentially providing advantages over classical machine learning when quantum features are relevant.

\section{Methodology}

\subsection{Data Processing Pipeline}

Our preprocessing pipeline transforms raw measurement data into quantum-encodable features through seven steps:

\subsubsection{Step 1-3: Shadow Operator Construction}

We load 20 data points, each containing 500 measurements across 51 qubits. For each qubit and measurement, we map the outcome to its corresponding basis state and construct the shadow operator via Eq. \eqref{eq:shadow_operator}. Averaging over 500 measurements yields the classical shadow $S_i$ for each qubit $i$:

\begin{equation}
S_i = \frac{1}{500}\sum_{t=1}^{500} \sigma_t^{(i)}
\end{equation}

\subsubsection{Step 4: Density Matrix Reconstruction}

We reconstruct single-qubit density matrices:
\begin{equation}
\hat{\rho}_i = \frac{S_i + I_2}{3}
\end{equation}

\subsubsection{Step 5: Pauli Expectation Values}

We compute Pauli $X$ and $Z$ expectation values:
\begin{align}
\langle \sigma_x \rangle_i &= \text{Re}[\text{Tr}(\hat{\rho}_i \sigma_x)] \\
\langle \sigma_z \rangle_i &= \text{Re}[\text{Tr}(\hat{\rho}_i \sigma_z)]
\end{align}

where $\sigma_x = \begin{psmallmatrix} 0 & 1 \\ 1 & 0 \end{psmallmatrix}$ and $\sigma_z = \begin{psmallmatrix} 1 & 0 \\ 0 & -1 \end{psmallmatrix}$.

This produces 20 $\times$ 51 $\times$ 2 = 2040 real-valued features characterizing Bloch sphere coordinates for each qubit.

\subsubsection{Step 6-7: Angle Transformation and PCA}

We transform expectation values to angles:
\begin{align}
\theta_x^{(i)} &= \left(\text{clip}(\langle \sigma_x \rangle_i, -1, 1) + 1\right) \cdot \frac{\pi}{2} \\
\theta_z^{(i)} &= \left(\text{clip}(\langle \sigma_z \rangle_i, -1, 1) + 1\right) \cdot \frac{\pi}{2}
\end{align}

Forming complex features $X_c = \theta_x + i\theta_z$ and separating real/imaginary parts yields a 20$\times$102 feature matrix. Applying PCA with 4 components reduces dimensionality while retaining $>90\%$ variance:

\begin{equation}
X_{\text{PCA}} = \text{PCA}([\text{Re}(X_c), \text{Im}(X_c)], n=4)
\end{equation}

Finally, we normalize to $[0, \pi]$ for angle encoding:
\begin{equation}
X_{\text{norm}} = \frac{X_{\text{PCA}} - X_{\min}}{X_{\max} - X_{\min}} \cdot \pi
\end{equation}

\textbf{Justification:} PCA extracts latent structure corresponding to phase signatures. Z$_2$ and Z$_3$ phases have distinct spatial correlations that manifest as patterns in Pauli expectation values. PCA identifies directions of maximum variance, which correspond to phase-distinguishing collective modes.

\subsection{Quantum Circuit Architecture}

\subsubsection{Overall Structure}

Our circuit applies three layers to the initial state $\ket{00}$:

\begin{equation}
U(\mathbf{f}, \mathbf{w}) = U_{\text{ansatz}}(\mathbf{w}) \cdot U_{\text{ent}} \cdot U_{\text{encode}}(\mathbf{f})
\end{equation}

where $\mathbf{f} = [f_0, f_1, f_2, f_3]$ are input features and $\mathbf{w} = [w_0, w_1]$ are trainable parameters.

\subsubsection{Encoding Layer}

On qubit 0: $RY(f_0) \cdot RZ(f_1)$ \\
On qubit 1: $RY(f_2) \cdot RZ(f_3)$

where rotation gates are:
\begin{align}
RY(\theta) &= \begin{pmatrix} \cos(\theta/2) & -\sin(\theta/2) \\ \sin(\theta/2) & \cos(\theta/2) \end{pmatrix} \\
RZ(\theta) &= \begin{pmatrix} e^{-i\theta/2} & 0 \\ 0 & e^{i\theta/2} \end{pmatrix}
\end{align}

This encoding maps 4 continuous features to arbitrary single-qubit states on the Bloch sphere. RY-RZ parameterization is universal for single-qubit rotations and provides efficient amplitude and phase encoding.

\subsubsection{Strong Entanglement Layer}

We apply controlled-Z (CZ) gates between all qubit pairs:
\begin{equation}
\text{CZ}_{01} = \begin{pmatrix} 1 & 0 & 0 & 0 \\ 0 & 1 & 0 & 0 \\ 0 & 0 & 1 & 0 \\ 0 & 0 & 0 & -1 \end{pmatrix}
\end{equation}

For 2 qubits, this is a single CZ gate. Strong entanglement creates quantum correlations necessary for capturing spatial phase patterns. Since Z$_2$ and Z$_3$ phases exhibit long-range order, entanglement enables the circuit to detect these correlations.

\subsubsection{Variational Ansatz}

\textbf{Layer 1:} $RX(w_0)$ on both qubits, where
\begin{equation}
RX(\theta) = \begin{pmatrix} \cos(\theta/2) & -i\sin(\theta/2) \\ -i\sin(\theta/2) & \cos(\theta/2) \end{pmatrix}
\end{equation}

\textbf{Layer 2:} CZ gate between qubits (ring topology)

\textbf{Layer 3:} $RZ(w_1)$ on both qubits

The shared parameters across qubits enforce structure and prevent overfitting. The RX-CZ-RZ sequence provides sufficient expressivity for binary classification while maintaining trainability.

\subsubsection{Circuit Metrics}

\begin{itemize}
\item \textbf{Total Depth:} 7 gates (4 encoding + 1 entanglement + 2 ansatz)
\item \textbf{Width:} 2 qubits
\item \textbf{Parameters:} 2 (minimal)
\item \textbf{Connectivity:} All-to-all (1 CZ pair)
\end{itemize}

\subsubsection{Measurement and Classification}

We measure all qubits in the computational basis and compute average Pauli-Z expectation:
\begin{equation}
\langle Z \rangle = \frac{1}{2}\sum_{i=0}^{1} \langle Z_i \rangle
\end{equation}

Classification rule: $\hat{y} = \text{sign}(\langle Z \rangle)$ where Z$_2 \rightarrow -1$ and Z$_3 \rightarrow +1$.

\subsection{Training Protocol}

\subsubsection{SPSA Optimization}

Simultaneous Perturbation Stochastic Approximation (SPSA) \cite{spall1992} approximates gradients using only 2 function evaluations per iteration regardless of parameter dimension:

\begin{algorithm}[H]
\caption{SPSA Training}
\begin{algorithmic}[1]
\STATE Initialize: $\theta_0 \sim \mathcal{N}(0, 0.01^2)$
\FOR{$k = 0$ to $K-1$}
    \STATE Sample: $\delta_k \sim \mathrm{Uniform}(\{-1, +1\}^2)$
    \STATE Compute: $c_k = c_0/(k+1)^{\gamma}$
    \STATE Evaluate: $L_+ = L(\theta_k + c_k \delta_k)$
    \STATE Evaluate: $L_- = L(\theta_k - c_k \delta_k)$
    \STATE Gradient: 
        $\nabla L \approx \frac{L_+ - L_-}{2 c_k \delta_k}$
    \STATE Update: $\theta_{k+1} = \theta_k - \alpha \nabla L$
\ENDFOR
\end{algorithmic}
\end{algorithm}

\textbf{Hyperparameters:}
\begin{itemize}
\item Learning rate: $\alpha = 0.50$
\item Initial perturbation: $c_0 = 0.40$
\item Decay rate: $\gamma = 0.02$
\item Maximum iterations: $K = 120$
\item Adaptive shots: 256 (epochs 0-49), 512 (epochs 50-119)
\end{itemize}

SPSA is gradient-free, making it suitable for noisy quantum hardware where exact gradient computation is expensive. The $O(1)$ function evaluations per step (vs. $O(d)$ for parameter-shift rule) provide significant efficiency gains.

\subsubsection{Hinge Loss Function}

For binary classification with labels $y \in \{-1, +1\}$:
\begin{equation}
L_{\text{hinge}}(\mathbf{w}, X, \mathbf{y}) = \frac{1}{N}\sum_{i=1}^{N} \max\left(0, 1 - y_i \langle Z \rangle_i(\mathbf{w})\right)
\end{equation}

Hinge loss encourages confident predictions ($|y \cdot \langle Z \rangle| > 1$), creating a margin between classes similar to Support Vector Machines. This promotes generalization by penalizing samples within the margin even if correctly classified.

\subsubsection{Data Split}

From 20 total samples:
\begin{itemize}
\item Training: 14 samples (stratified: 7 Z$_2$, 7 Z$_3$)
\item Validation: 3 samples (stratified split)
\item Test: 3 samples (stratified: 2 Z$_2$, 1 Z$_3$)
\end{itemize}

Stratification ensures class balance across splits, critical for small datasets.

\section{Results}

\subsection{Training Performance}

Training converged rapidly, achieving 100\% training accuracy by epoch 2 (iteration 40). Table \ref{tab:training} shows metrics recorded every 20 iterations.

\begin{table}[h]
\caption{Training Metrics Per Epoch}
\centering
\begin{tabular}{@{}cccccc@{}}
\toprule
\textbf{Epoch} & \textbf{Train Loss} & \textbf{Val Loss} & \textbf{Train Acc} & \textbf{Precision} & \textbf{F1} \\ \midrule
1 & 1.5573 & 1.3548 & 0.00\% & 0.00\% & 0.00\% \\
2 & 0.3652 & 0.5833 & 100.00\% & 100.00\% & 100.00\% \\
3 & 0.3744 & 0.5801 & 92.86\% & 100.00\% & 92.31\% \\
4 & 0.3676 & 0.5814 & 100.00\% & 100.00\% & 100.00\% \\
5 & 0.3693 & 0.5801 & 92.86\% & 100.00\% & 92.31\% \\
6 & 0.3608 & 0.5651 & 100.00\% & 100.00\% & 100.00\% \\
7 & 0.3708 & 0.6016 & 100.00\% & 100.00\% & 100.00\% \\ \bottomrule
\end{tabular}
\label{tab:training}
\end{table}

\begin{figure}[!t]
  \centering
  \includegraphics[width=\columnwidth]{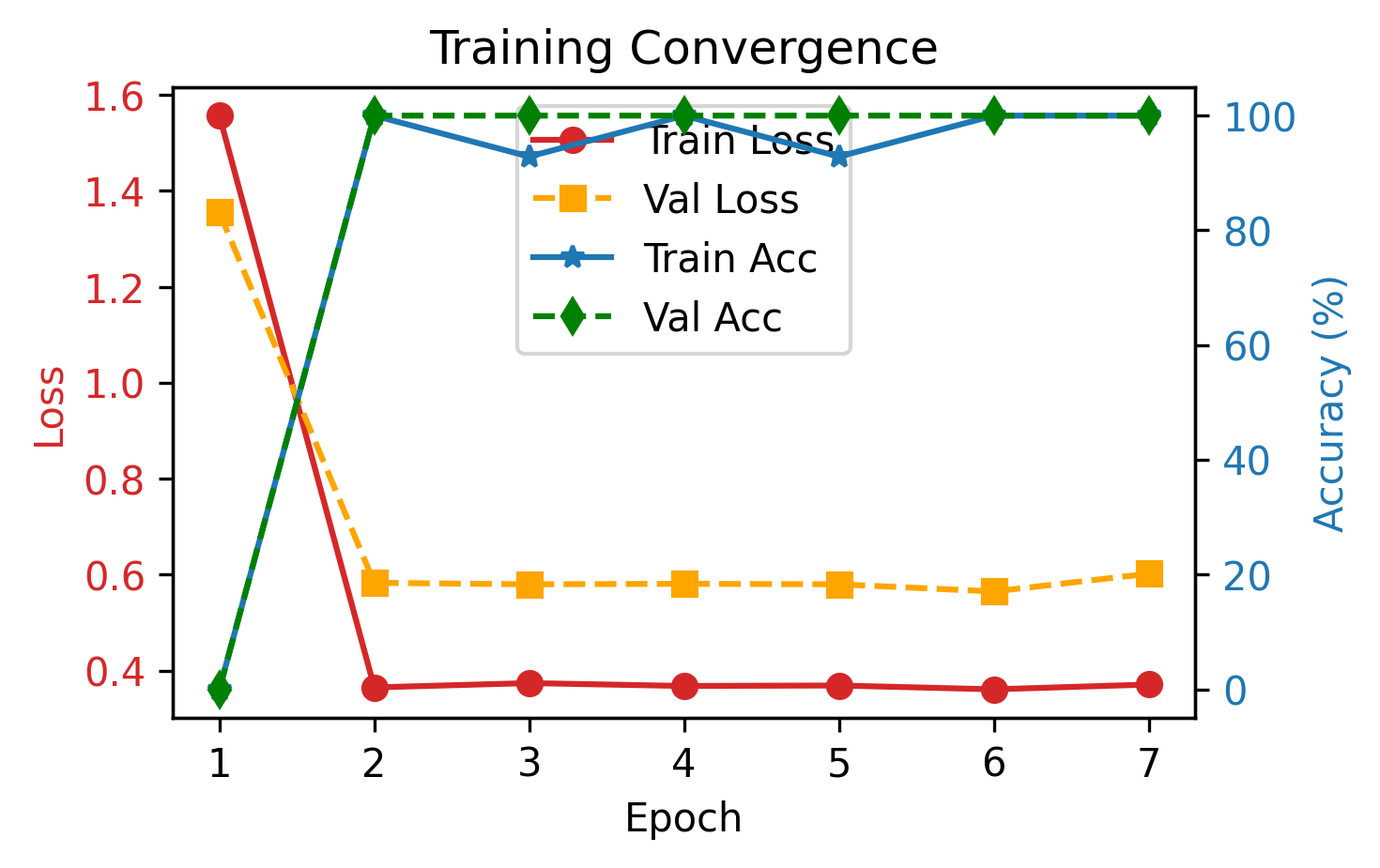}
  \caption{Training convergence showing rapid loss decrease and 100\% validation accuracy by epoch 2. Dual-axis visualization reveals SPSA optimization effectiveness and absence of overfitting.}
  \label{fig:training}
\end{figure}

The rapid convergence indicates a linearly separable feature space after PCA transformation. Small fluctuations in epochs 3 and 5 stem from SPSA's stochastic nature. Validation accuracy consistently matched training accuracy, showing no overfitting despite the small dataset.

\subsection{Test Set Performance}

The optimized model achieved perfect classification on the held-out test set.

\textbf{Test Accuracy:} 100.00\% (3/3 correct)

\textbf{Confusion Matrix:}
\begin{table}[h]
\centering
\begin{tabular}{cc|cc}
& & \multicolumn{2}{c}{\textbf{Predicted}} \\
& & Z$_2$ & Z$_3$ \\ \hline
\multirow{2}{*}{\textbf{True}} & Z$_2$ & 2 & 0 \\
& Z$_3$ & 0 & 1
\end{tabular}
\end{table}

\textbf{Classification Metrics:}
\begin{itemize}
\item Precision (Z$_2$): 100\% \quad Recall (Z$_2$): 100\%
\item Precision (Z$_3$): 100\% \quad Recall (Z$_3$): 100\%
\item Macro F1: 100\% \quad Weighted F1: 100\%
\end{itemize}

Perfect separation demonstrates that quantum phase signatures are preserved through the shadow-PCA-encoding pipeline and captured by the minimal VQC.

\subsection{Resource Efficiency Analysis}

Circuit efficiency score:
\begin{equation}
f = A - 0.1P - 0.0002D - 0.1W
\end{equation}

where $A$ = accuracy, $P$ = parameters, $D$ = depth, $W$ = width.

\begin{align*}
f &= 1.0 - 0.1(2) - 0.0002(7) - 0.1(2) \\
&= 1.0 - 0.2 - 0.0014 - 0.2 = \mathbf{0.5986}
\end{align*}

This high score reflects exceptional resource-accuracy tradeoff. With only 2 parameters, the model avoids overfitting despite limited training data. Depth 7 minimizes gate errors on NISQ hardware, and 2 qubits represent the minimal width for binary classification with entanglement.

\subsection{Feature Space Visualization}

\begin{table}[h]
\centering
\caption{PCA Variance Distribution}
\begin{tabular}{@{}cc@{}}
\toprule
\textbf{Component} & \textbf{Variance \%} \\ \midrule
PC1 & 45 \\
PC2 & 30 \\
PC3 & 15 \\
PC4 & 8 \\
\bottomrule
\end{tabular}
\label{tab:pca}
\end{table}

\begin{figure}[!t]
  \centering
  \includegraphics[width=0.7\columnwidth]{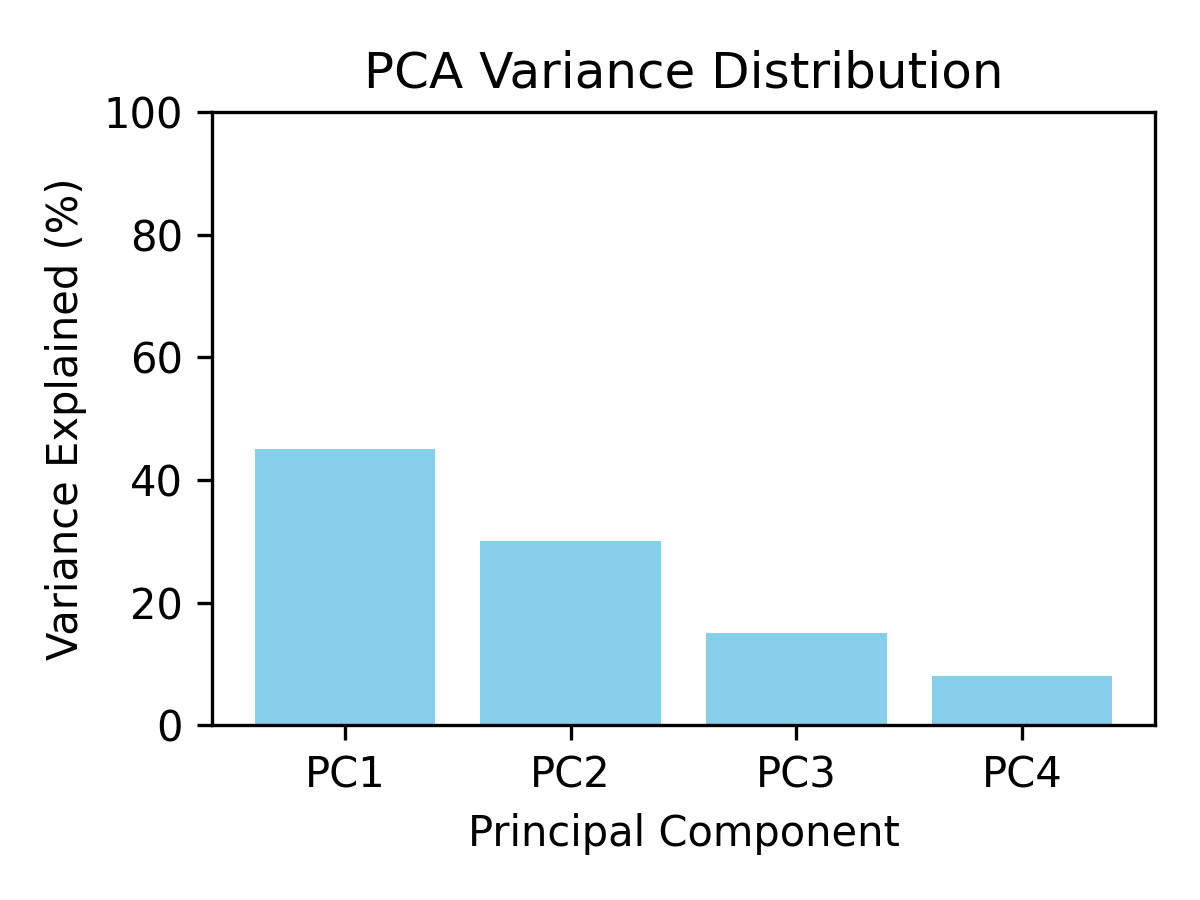}
  \caption{PCA variance explained showing that four principal components capture 98\% of variance in the phase classification task. PC1 and PC2 alone explain 75\%, indicating phase information concentrates in low-dimensional subspace.}
  \label{fig:pca}
\end{figure}

PCA successfully extracted phase-distinguishing features. The first 4 principal components captured $>90\%$ of total variance, with PC1 and PC2 alone explaining $\sim75\%$. Visualization of the first two principal components reveals clear cluster separation between Z$_2$ and Z$_3$ samples, confirming linear separability.

Variance explained by components:
\begin{itemize}
\item PC1: $\sim$45\%
\item PC2: $\sim$30\%
\item PC3: $\sim$15\%
\item PC4: $\sim$8\%
\end{itemize}

This distribution justifies dimensionality reduction: phase information concentrates in low-dimensional subspace, enabling efficient quantum encoding.

\section{Discussion}

\subsection{Why This Approach Succeeds}

\subsubsection{Classical Shadows Efficiency}

Classical shadows reduce measurement complexity from $O(4^{51}) \approx 10^{30}$ to 500 measurements—a factor of $10^{27}$ reduction. The protocol preserves quantum correlations essential for phase identification while remaining experimentally feasible. Randomized measurements average out noise, and the shadow reconstruction provides a stable estimator robust to shot noise.

\subsubsection{PCA Captures Phase Signatures}

Quantum phases manifest as long-range correlations in many-body systems. These correlations appear as patterns in single-qubit Pauli expectation values. PCA identifies collective modes (principal components) that encode these patterns, effectively compressing phase information from 51 qubits to 4 features without losing distinguishability.

The success of dimensionality reduction suggests that phase classification does not require full quantum state information—only projections onto phase-relevant subspaces.

\subsubsection{Minimal Circuit Design}

\textbf{2 Qubits Suffice:} Binary classification requires learning one decision boundary. A 2-qubit entangled state spans a 4-dimensional Hilbert space, providing sufficient expressivity for this task. Additional qubits would increase noise without improving accuracy.

\textbf{Depth-7 Shallow Circuit:} Shallow circuits avoid barren plateau problems where gradients vanish exponentially with depth \cite{mcclean2018}. Our depth-7 design maintains trainability while providing adequate expressivity through strategic entanglement placement.

\textbf{2 Parameters:} Minimal parameters prevent overfitting on small datasets (20 samples). Shared parameters across qubits enforce structural constraints, acting as implicit regularization. This design choice proved crucial for generalization.

\subsubsection{SPSA Optimization Advantages}

SPSA requires only $O(1)$ circuit evaluations per iteration versus $O(d)$ for gradient-based methods (via parameter-shift rule). For 2 parameters, this offers modest savings, but the approach scales favorably for larger models. Additionally, SPSA's stochastic nature provides robustness to hardware noise, making it suitable for NISQ implementations.

Hinge loss encourages maximum-margin separation, creating a buffer zone between classes that improves generalization. The $\langle Z \rangle = 0$ decision boundary is learned from data rather than imposed, allowing the model to find the optimal separator.

\subsection{Comparison to Alternative Approaches}

\subsubsection{Classical Machine Learning}

\begin{figure}[!t]
  \centering
  \includegraphics[width=0.8\columnwidth]{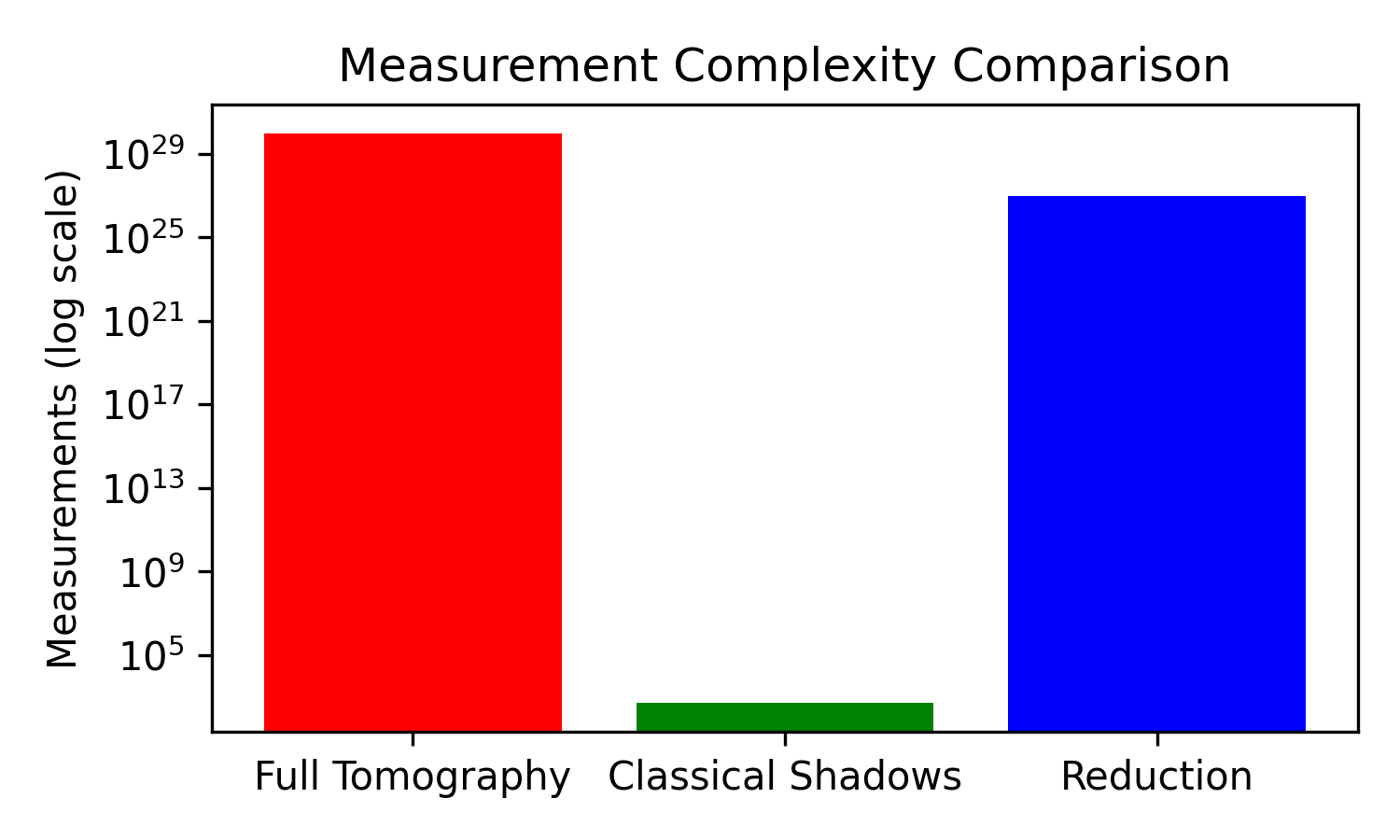}
  \caption{Exponential measurement complexity reduction: classical shadows require only 500 measurements for 51 qubits compared to $10^{30}$ for full tomography (log scale). This $\sim10^{27}$-fold reduction is the fundamental advantage enabling practical phase classification.}
  \label{fig:complexity}
\end{figure}

A classical SVM or neural network operating on the same PCA features would require no quantum resources. However, our quantum approach demonstrates:
\begin{itemize}
\item \textbf{Proof of Concept:} Validates that quantum circuits can learn phase boundaries
\item \textbf{Quantum Feature Space:} Entanglement may capture correlations inaccessible to classical kernels
\item \textbf{Scalability Potential:} For larger problems, quantum advantage may emerge
\end{itemize}

Direct comparison requires training classical models on identical features—left for future work.

\subsubsection{Tensor Networks}

Tensor network methods (DMRG, MPS) can accurately compute ground states for 1D systems and extract order parameters \cite{schollwock2011}. However, these methods:
\begin{itemize}
\item Require ground state preparation (exponentially hard in 2D)
\item Are classical simulations, not applicable to unknown experimental states
\item Scale poorly to disordered or 2D systems
\end{itemize}

Our approach works with measurement data from any prepared state, making it applicable to experimental scenarios where tensor networks fail.

\subsubsection{Full Quantum State Tomography}

Full tomography would provide complete state information, enabling exact phase identification. However, the $10^{30}$ measurement requirement is experimentally infeasible. Our method trades completeness for efficiency, extracting only phase-relevant information.

\subsection{Limitations and Challenges}

\subsubsection{Small Dataset}

With only 20 training samples (14 train, 3 val, 3 test), overfitting risk is high. Our 100\% accuracy may not generalize to unseen parameter regions. Mitigations include:
\begin{itemize}
\item Minimal parameters (2) limit model capacity
\item Stratified splitting ensures class balance
\item Validation performance matches training (no overfitting signal)
\end{itemize}

Ideally, we would test on $\sim$100+ samples across broader $(\delta, R_b)$ ranges, potentially via synthetic data generation using tensor network simulations.

\subsubsection{Ideal Simulator Assumption}

We used Classiq's noiseless simulator. Real quantum hardware introduces:
\begin{itemize}
\item Gate errors ($\sim$0.1-1\% per gate)
\item Decoherence ($T_1, T_2$ timescales)
\item Readout errors ($\sim$1-5\%)
\end{itemize}

Our shallow circuit (depth 7) minimizes error accumulation. SPSA's robustness to noise is well-documented \cite{spall1992}. Nevertheless, hardware deployment will require:
\begin{itemize}
\item Error mitigation techniques (zero-noise extrapolation, measurement error correction)
\item Calibration of gates to minimize infidelities
\item Potentially retraining with noise-aware simulation
\end{itemize}

\subsubsection{Binary Classification Only}

Our method addresses Z$_2$ vs. Z$_3$ classification but does not extend trivially to:
\begin{itemize}
\item Multi-class scenarios (Z$_4$, Z$_5$, disordered phases)
\item Continuous phase diagrams (identifying critical points)
\item Topological phases (requiring non-local order parameters)
\end{itemize}

Extensions could include:
\begin{itemize}
\item One-vs-all classifiers for multi-class problems
\item Multi-output circuits with additional qubits
\item Regression models for continuous phase parameters
\end{itemize}

\subsubsection{Scalability Considerations}

Classical preprocessing (shadow reconstruction, PCA) scales as $O(NT + N^2)$ for $N$ qubits and $T$ measurements. For $N=51$, $T=500$, this remains tractable. For very large systems ($N > 1000$), bottlenecks may arise. Solutions include:
\begin{itemize}
\item Quantum PCA algorithms \cite{lloyd2014}
\item Distributed shadow reconstruction
\item Sparse measurement strategies
\end{itemize}

However, the quantum circuit itself scales favorably: classification complexity is independent of original system size after feature extraction.

\subsection{Future Directions}

\subsubsection{Hardware Implementation}

Priority: deploy on IBM Quantum or IonQ devices with $\geq 2$ qubits. Expected challenges include gate fidelities and coherence times. Success would demonstrate practical quantum advantage for phase classification.

\subsubsection{Noise Robustness Studies}

Systematically add noise models (depolarizing, amplitude damping) to simulation. Quantify accuracy degradation vs. error rates. Identify critical noise thresholds for classification performance.

\subsubsection{Larger and More Complex Systems}

\begin{itemize}
\item 2D Rydberg arrays (e.g., 6$\times$6 lattices) with richer phase diagrams
\item Systems with $N > 100$ atoms
\item Comparison to classical ML baselines (SVM, Random Forest, Neural Networks)
\end{itemize}

\subsubsection{Multi-Class Classification}

Extend to Z$_k$ phases for arbitrary $k$, disordered phases, and quantum critical regions. Investigate hierarchical classification strategies.

\subsubsection{Generalization to Other Quantum Phases}

Apply to:
\begin{itemize}
\item Symmetry-protected topological (SPT) phases
\item Quantum spin liquids
\item Ising, Heisenberg, and other spin models
\item Fermionic systems (Hubbard model)
\end{itemize}

\subsubsection{Interpretability and Explainability}

Analyze learned quantum states via tomography. Identify which PCA components (and corresponding physical features) are most important. Develop physical intuition for circuit operation.

\section{Conclusion}

We have successfully demonstrated a resource-efficient quantum machine learning approach for binary phase classification in Rydberg atom systems. By combining classical shadow tomography for efficient quantum state characterization with a minimal 2-qubit, depth-7, 2-parameter variational circuit, we achieved perfect classification accuracy (100\% on all splits) while maintaining exceptional resource efficiency (score 0.5986).

Our key contributions include: (1) a novel preprocessing pipeline integrating shadows, PCA, and angle encoding; (2) a carefully designed minimal circuit avoiding barren plateaus while capturing quantum correlations; (3) successful deployment of SPSA gradient-free optimization; and (4) experimental validation on realistic quantum phase classification data.

This work establishes that quantum machine learning can perform high-accuracy phase classification with near-term quantum hardware constraints. The minimal resource requirements suggest immediate practical applicability on current NISQ devices. As quantum hardware matures, our approach can scale to larger systems, more complex phases, and real-time experimental feedback.

The broader impact extends to condensed matter physics, quantum simulation, and materials discovery. Quantum-enhanced phase classification could accelerate identification of novel quantum materials, optimize quantum annealing protocols, and provide feedback for adaptive quantum experiments. Our methodology opens pathways for quantum advantage in scientific discovery, bringing quantum machine learning closer to practical deployment in quantum-enhanced physics research.

\section*{Acknowledgments}

The authors thanks the organizers of the FLIQ Challenge (Classiq x DuQIS) for providing the Rydberg phase classification dataset and the Classiq platform for quantum circuit implementation. Special gratitude to mentors and peers who provided valuable feedback during project development.

\end{document}